\documentclass[nonacm]{acmart}

\usepackage[english]{babel}

\usepackage{amsmath}
\usepackage{graphicx}
\usepackage{wrapfig}
\usepackage{listings}

\usepackage{enumitem}

\usepackage{xcolor}
\usepackage{xparse}

\title{KwicKwocKwac, a tool for rapidly generating concordances and marking up a literary text}
\author{Sebastian Barzaghi}
\email{sebastian.barzaghi2@unibo.it}
\orcid{0000-0002-0799-1527}
\affiliation{%
  \institution{DBC - University of Bologna}
  \city{}
  \country{Italy}
}

\author{Francesco Paolucci}
\email{francesco.paolucci7@unibo.it}
\affiliation{%
  \institution{DISI - University of Bologna}
  \city{}
  \country{Italy}
}

\author{Francesca Tomasi}
\affiliation{%
  \institution{FICLIT - University of Bologna}
  \city{}
  \country{Italy}
}
\email{francesca.tomasi@unibo.it}

\author{Fabio Vitali}
\affiliation{%
  \institution{DISI - University of Bologna}
  \city{}
  \country{Italy}
}
\email{fabio.vitali@unibo.it}

\begin{document}

\begin{abstract}
This paper introduces KwicKwocKwac 1.0 (KwicKK), a web application designed to enhance the annotation and enrichment of digital texts in the humanities. KwicKK provides a user-friendly interface that enables scholars and researchers to perform semi-automatic markup of textual documents, facilitating the identification of relevant entities such as people, organizations, and locations. Key functionalities include visualization of annotated texts using KeyWord in Context (KWIC), KeyWord Out Of Context (KWOC), and KeyWord After Context (KWAC) methodologies, alongside automatic disambiguation of generic references and integration with Wikidata for Linked Open Data connections. The application supports metadata input and offers multiple download formats, promoting accessibility and ease of use. Developed primarily for the National Edition of Aldo Moro’s works, KwicKK aims to lower the technical barriers for users while fostering deeper engagement with digital scholarly resources. The architecture leverages contemporary web technologies, ensuring scalability and reliability. Future developments will explore user experience enhancements, collaborative features, and integration of additional data sources.
\end{abstract}

\maketitle

\pagestyle{empty}

\section{Introduction}

KwicKwocKwac 1.0 (KwicKK) is a Web application that provides scholars and researchers with a simple and intuitive tool to enrich the text of documents in digital format through semi-automatic markup and metadata features and the visualization of interesting features in the documents marked up with it, with particular attention to concordances according to the KWIC methodology.

The main features offered by KwicKK are:
\begin{enumerate}
    \item Marking of relevant entities within the text of the document (mentions to people, organizations, places, bibliographical references and citations);
    \item Visualization of text mentions in a grouped and sorted interface according to the approaches known as KeyWord in Context (KWIC), KeyWord Out Of Context (KWOC) and KeyWord After Context (KWAC); 
    \item Automatic recognition of the types of notes in the text (researcher's notes and Moro notes);
    \item Disambiguation of generic references to entities to which the document refers (e.g., terms such as "President of the Council," "Pope," etc.);
    \item Reconciliation with data on Wikidata to disambiguate tagged entities and connect them to the Web in a Linked Open Data (LOD) perspective;
    \item Input of bibliographic metadata related to the document (e.g., document topic, document type, etc.);
    \item Downloading the marked document in multiple formats (e.g., HTML and TEI-XML).
\end{enumerate}

The environment, while user-friendly, is accompanied by comprehensive and timely documentation, including the guidelines, instructions aimed at preparing volumes for their optimal semi-automatic processing, the handbook, instructions intended to provide a comprehensive overview of KwicKK's functionality to guide users as they use it, and the online documentation, which consists of a list of the main functions of KwicKK accompanied by video tutorials to easily find the information needed to perform the most important operations.

The tool was mainly adopted within the National Edition of Aldo Moro's works\footnote{\url{https://aldomorodigitale.unibo.it/}}, a digital scholarly edition that was developed by the Digital Humanities Advanced Research Centre (/DH.arc) at the University of Bologna.  encompassing both published and unpublished texts of an important historical, cultural and political figure in the Italian landscape towards the end and after the Second World War. The edition -- with its 484 fully digitized documents, enriched with bibliographic metadata encoded in RDF, identified with DOIs, and available for download in various formats, including RDFa-HTML and TEI-XML -- intends to be a web-based scientific resource for experts and the public alike to delve into Aldo Moro's life and work through exploration, visualization, and download of a data-rich corpus of texts. In this context, KwicKK was used to semi-automatically annotate the entities (people, places, organizations, bibliographic references and quotations) in the texts and associate a set of descriptive metadata to the documents.

\section{Background and related work}

\subsection{Keyword-based subject indexing systems}

KeyWord In Context (KWIC) is an automatic subject indexing system that focuses on the titles of textual resources. Initially proposed as "Keywords in Titles" by Crestadoro in 1856, the concept was revisited by Luhn in 1958 during his work at IBM \cite{williamshans2010}. KWIC operates on the premise that a document's title serves as a concise abstract of its content. By identifying significant terms — excluding stopwords and other non-essential words — KWIC enables effective subject indexing of documents \cite{luhnkey1960}. Another important feature of KWIC is preserving the context around each keyword in the index entries. Each entry is generated by selecting a term as the main keyword, along with its neighbors in the title. This process causes the title to shift sideways, positioning the keyword either on the far left or in the center of the line. Subsequently, all entries are organized in alphabetical order.

Over the following years, variations of KWIC have come about to enhance retrieval efficiency and improve entry presentation. Notable variants include KeyWord Out of Context (KWOC) and KeyWord Alongside Context (KWAC). In KWOC, the entry keyword is positioned on the left, followed by the entire title to provide the complete, readable context. Meanwhile, KWAC - much like KWIC - shifts the keywords to the left for emphasis, but retains the surrounding context, reflecting how they are actually used.

A literature search reveals that KWIC, and its derivates, despite -- or confirming -- their longevity, have been and are still used as useful tools for indexing textual information. For example, \cite{kaki2006fkwic} proposes a tool to filter Web search results based on the frequency of keyword contexts. When a context is selected, the relevant results are displayed, enabling users to focus on specific information. In an experiment involving 36 participants, the KWIC-based user interface was compared with a traditional rank-order result listing. The findings revealed that the proposed interface was faster and more precise in retrieving relevant result.

In another study (\cite{san2013kwic}), a pilot investigation assessed the efficacy of a specialized term list -- generated through a term extraction process applied to a corpus of KWIC entries -- as a source of definitional information. The foundational hypothesis posited that the high frequency of a lexical unit within the KWIC concordance lines of a term serves as an indicator of its potential relevance in the term's definition. The results suggest that a KWIC corpus, when combined with a generated term list, could be an effective tool for formulating definitions.

The study conducted by (\cite{safii2021index}) aims to leverage KWIC to create an index for the online catalog of the Batu City Public Library. The research utilizes bibliographic data originally encoded in MARC-XML and collected via the Open Archives Initiative Protocol for Metadata Harvesting (OAI-PMH) from the library's server. After a phase of cleaning and normalization, the KWIC index is created and tested. The findings indicate that KWIC enhances the users' ability to retrieve information effectively from the online catalog of the library.

\subsection{Scholarly editing tools}
In recent years, various software tools and services have emerged to help humanities scholars transcribe, edit and annotate texts with metadata and markup (such as XML-TEI and RDF) in order to produce scholarly digital editions to be published on the Web \cite{spadiniediting2015}. However, a key challenge with many of these tools is the technical expertise required to configure and use them, which often prevents researchers from focusing on their scientific work.

Given the technical barrier represented by text encoding, many of these tools focus mainly on editing. This entails modifying the text and adding structural, contextual, and content-related annotations to it. For example, the Canadian Writing Research Collaboratory developed CWRC-Writer, a web-based text markup editor, for collaborative scholarly editing projects \cite{browncwrc-writer2014}. CWRC-Writer aims to facilitate the production of semantically-enriched documents by scholars with limited knowledge of RDF and XML markup, while still allowing the more technology-oriented users to work on the XML source code that constitutes the document. It combines elements such as XML-TEI markup for structure and content (e.g. structural tags and entities) with RDF for representing standoff annotations and the relations between elements. Another interesting feature of this tool is the reconciliation functionality that allows one to align annotated entities with records existing in controlled vocabularies (e.g. VIAF or GeoNames) or defined in internal authority lists. However, CWRC-Writer does present certain limitations, including the lack of facsimile management, and a user-friendly mechanism to inject descriptive metadata into the document itself.

Another notable project is the Standoff Property Editor (SPEEDy) \cite{neill2021speedy}, a web-based application that enables users to transcribe, edit, and annotate texts. It employs HTML markup tags and properties, allowing for straightforward updates to content, formatting, and annotations all in one place. The application is designed to separate the text from the added properties and annotations, avoiding direct embedding into the text stream. Instead, it references the start and end character indexes, ensuring that the text remains mutable; as modifications occur, the standoff properties associated with the annotated text are updated accordingly. However, it does have limitations, such as the inability to import or export XML-TEI documents, and challenges in handling multi-layered texts of larger sizes.

More recently, a web application called Śivadharma Database \cite{tomasi2024sivadharma} is being developed at the /DH.arc Research Centre of the Department of Classical Philology and Italian Studies at the University of Bologna. It is aimed at streamlining the creation of scholarly digital editions by including their storage, publication, cataloging, updating, visualization and browsing. The tool allows scholars to work on the text and its components, add descriptive metadata, manage the critical apparatus as well as other structures such as notes, transcriptions, translations and citations, with the aim to ensure consistent scholarly data representation and management, simplify the publishing process, and allow outputs to be easily exported as XML-TEI documents. 

While existing web-based annotation tools like CWRC-Writer, SPEEDy, and Śivadharma Database make significant strides in enhancing the capabilities of humanities scholars to annotate and manage texts, they often fall short in addressing certain functionalities, such as metadata addition and manipulation of marked-up entities. We aim to add KwicKK in the landscape of existing markup tools to further empower scholars to engage more deeply with their research without the steep learning curve associated with traditional markup languages.

\section{Architecture of KwicKK}

KwicKK runs on a Node.js server and was developed with current web technologies. It uses the Boostrap library for styling and layout, and several Javascript frameworks to provide a more convenient HTML document manipulation. 

The application core libraries are:

\begin{itemize}
    \item \textit{script.js}: the main function that sets up the markup environment;
    \item \textit{kwic.js}: starting from the creation of a simple \verb|Range| DOM object, enriches the HTML document with RDFa vocabulary. Handles the selected nodes and initiates the annotation process defining a nested structure of mentions, entities and categories. Dealing with any changes related to that;
    \item \textit{util.js}: a set of functions used across the entire application that extend \verb|Range| and \verb|Node| prototypes;
    \item \textit{auth.js}: user authentication library and ecrypted state management via JSON Web Token (JWT).
\end{itemize}

The full source code of KwicKK is available on Github\footnote{\url{https://github.com/sanofrank/KwicKwocKwac}}.

\section{Interface of KwicKK}

\begin{figure}[ht]
  \centering
  \includegraphics[width=\linewidth]{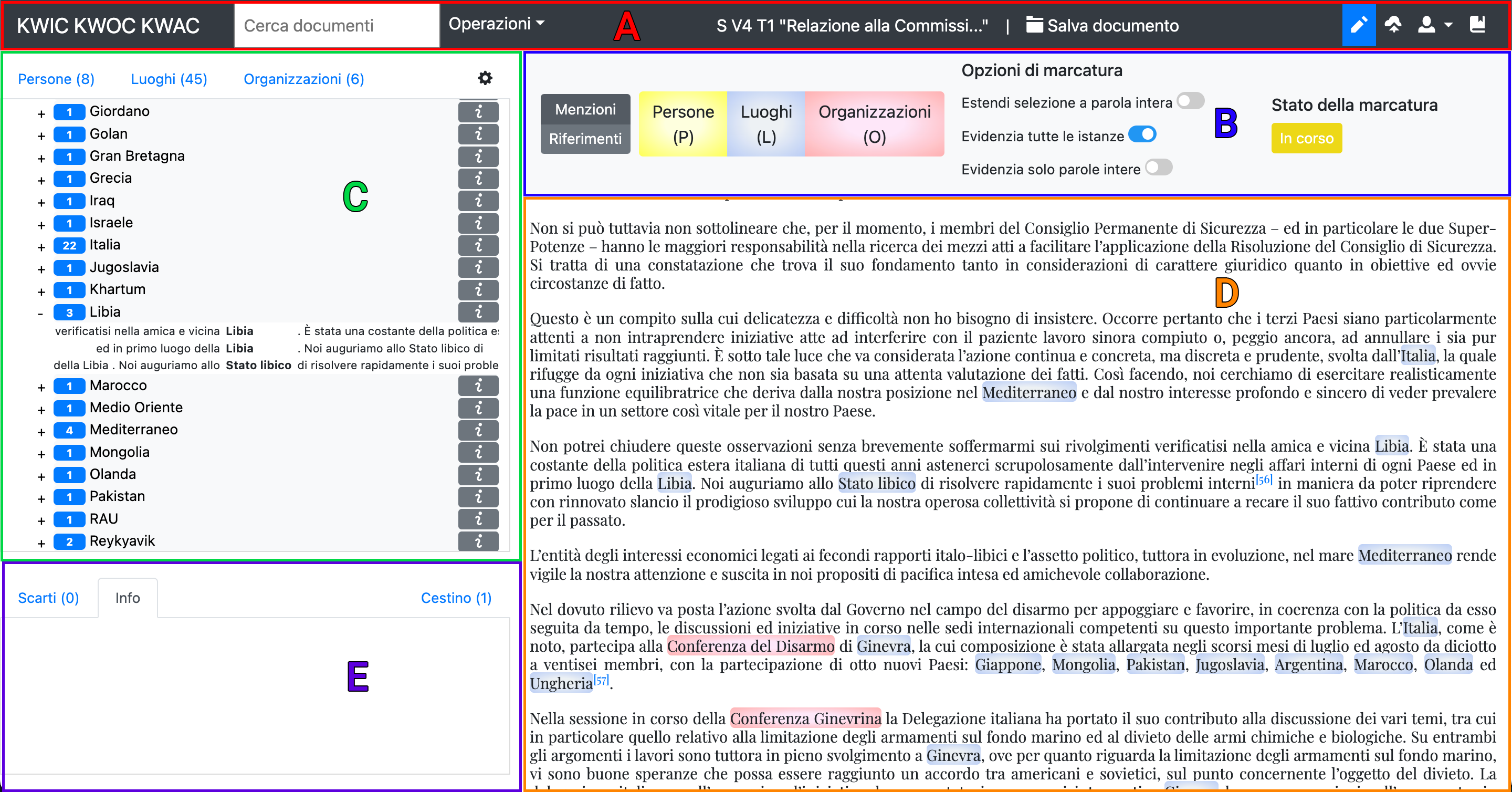}
  \caption{The interface of KwicKK, divided into five main sections: A) navigation bar; B) toolbar; C) entity panel; D) document area; E) additional utilities.}
  \Description{The interface of KwicKK, divided into five main sections: A) navigation bar; B) toolbar; C) entity panel; D) document area; E) additional utilities.}
  \label{main}
\end{figure}

As illustrated in Figure \ref{main}, the interface is divided into five main areas, each explained in finer detail in the rest of the paper:
\begin{enumerate}[label=\Alph*]
    \item The navigation bar (Section \ref{sec:navigation}), which allows the user to navigate the documents and do other operations on the document she is working on, such as adding metadata and saving it;
    \item The toolbar (Section \ref{sec:toolbar}), which contains the main functionalities add markup to the text of the document;
    \item The entity panel (Section \ref{sec:entitypanel}), which stores and visualizes the list of entities marked up in the text, and allows the user to manipulate them in varius ways;
    \item The document area( Section \ref{sec:documentarea}), where both the actual text of the document and the mentions highlighted by the user are visualized;
    \item Additional utilities (Section \ref{sec:additionalutilities}), which provides the user with functionalities to work with scrapped mentions and to visualize information related to entities if they are linked successfully to a Wikidata record.
\end{enumerate}

\subsection{The navigation bar}
\label{sec:navigation}
The navigation bar (see area A in Figure \ref{main}) has the following features:
\begin{enumerate}
    \item Document search: search bar that lists uploaded documents and allows filtered searches and selections;
    \item Operations menu: allowing users to activate/deactivate markup display, add metadata, download documents in TEI or HTML format, export/import identified entities, empty the recycle bin;
    \item Save document command to save the edited document;
    \item The loading of new documents into the marking environment;
    \item The activation/deactivation of the editing mode;
    \item the user profile menu where the user can change the password and exit the application;
    \item the access to the documentation, where the user can find basic instructions for using KwicKK;
    \item the access to the policy and copyright of the application;
\end{enumerate}

\subsection{The toolbar}
\label{sec:toolbar}

The toolbar (see area B in Figure \ref{main}) becomes visible and usable in editing mode. It has the following features:
\begin{enumerate}
    \item marking of selected text;
    \item Extend selection to whole word;
    \item Highlight all instances;
    \item Change marking status;
\end{enumerate}
See Section \ref{sec:editing} for further details.

\subsection{The entity panel}
\label{sec:entitypanel}
The entity panel (see area C in Figure \ref{main}) contains a series of horizontally arranged tabs. Each tab corresponds to a particular category of entities present in the text and highlighted by the user. Activating a tab by mouse click opens an index of all the elements marked in the text and belonging to that particular category, as discussed in Section \ref{sec:intratextual}.

\subsection{The document area}
\label{sec:documentarea}
The document area (see area D in Figure \ref{main}) contains the text of the document to be annotated. 

\subsection{Additional utilities}
\label{sec:additionalutilities}
The bottom left area (see area E in Figure \ref{main}) contains the following additional utilities, organized into panels that can be activated by clicking on their respective tabs:
\begin{enumerate}
    \item a Scrap panel, in which you can move items contained in the entity panel by dragging and dropping the selection, to hold them aside pending further operations;
    \item an Info panel, containing a body of informational text that automatically updates when the user clicks on an item synchronized with Wikidata (marked by a green checked checkbox);
    \item A Trash panel, in which you can move items contained in the panel of entities or in the Discard panel to delete them. The Trash panel behaves like the "Recycle Bin" feature of common operating systems with a graphical user interface. The user can empty it -- thus permanently deleting the items placed inside it -- by clicking on the Operations item in the navigation bar and selecting the Empty Trash  option in the drop-down menu.
\end{enumerate}

\section{Working with KwicKK}

\subsection{Marking up documents}
\label{sec:editing}
\begin{figure}[ht]
  \centering
  \includegraphics[width=\linewidth]{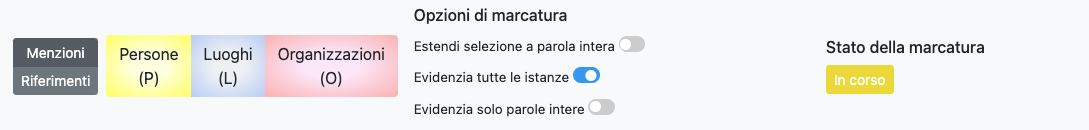}
  \caption{Markup toolbar}
  \Description{Markup toolbar}
  \label{toolbar}
\end{figure}

The toolbar (Figure \ref{toolbar}) controls the main features which allow the user to mark intertextual elements either as Mentions or References. On the one hand, Mentions are classified into these categories: People, Places or Organizations. On the other, References consists of Bibliographic references and Quotations. The list is totally parameterized and can be expanded and modified at will. It is also possible to automatically extend the selection to the entire word or highlight all occurrences that match the selected string in the entire text. 

Once a string is selected and one of the categories has been chosen, KwicKK wraps a \texttt{span} HTML element containing the following attributes:

\begin{lstlisting}
<span id="mention-1" typeof="<CLASS>" about="#mention-1" 
class="mention organization" property="dcterms:references"
resource="#DemocraziaCristiana">DC</span>
\end{lstlisting}

\begin{itemize}
    \item \verb|id|: the mention identification number;
    \item \verb|typeof|: the class used to represent the mention, as defined in an ontology; 
    \item \verb|about|: the mention reference that redirects the browser window to that mention;
    \item \verb|property|: the property the mention uses to reference the respective entity, as defined in the Dublin Core schema;
    \item \verb|resource|: the entity that the mention refers to.
\end{itemize}

Both the mentions and the entites they are related to -- listed with meta tags in the HTML document head -- follow the RDFa standard, a specification built on RDF that extends HTML so as to allow the addition of structured data directly into HTML markup using attributes.
  
\begin{lstlisting}
<meta about="#DemocraziaCristiana" typeof="<CLASS>">
<meta about="#DemocraziaCristiana" property="rdfs:label" 
content="Democrazia Cristiana">
<meta about="#DemocraziaCristiana" property="dcterms:relation"
resource="http://www.wikidata.org/entity/Q815348">
\end{lstlisting}

\begin{itemize}
    \item \verb|typeof|: the class used to represent the entity, as defined in an ontology;
    \item \verb|property="rdfs:label"|: a property reusing \verb|rdfs:label| from the RDF Schema vocabulary (RDFS)\footnote{\url{http://www.w3.org/2000/01/rdf-schema}} to provide a human-readable version of a resource's name;
    \item \verb|property="dcterms:relation"|: a property reusing \verb|dcterms:relation| from DCTerms to link the entity to an external resource (in this case, the Wikidata record related to the entity).
\end{itemize}

The document’s status is an internal feature that enables deeper filtering options to facilitate search and work management. It has three status options: "To be started" means the document has just been uploaded; "In progress" indicates the researcher is working on it; "Finished" shows that the work on the document is complete.

\subsection{Entities manipulation and disambiguation}
\label{sec:intratextual}
\begin{figure}[ht]
  \centering
  \includegraphics[width=\linewidth]{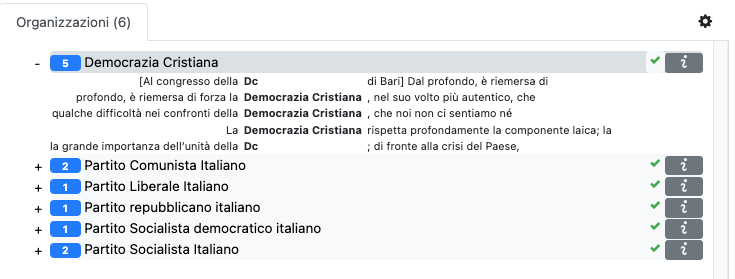}
  \caption{Mentions referring to the entity Democrazia Cristiana}
  \Description{Mentions referring to the entity Democrazia Cristiana}
  \label{entitypanel}
\end{figure}

Entities are divided in Category tabs inside the Entity panel (Figure \ref{entitypanel}). Clicking on one entity opens an index of all its intertextual mentions in the document. The mentions are showed by default as KWIC entries, but this can also be changed in the panel settings to appear as KWOC or KWAC entries. 

The user can interact in several ways with the mentions and references contained in the entity panel. For example, the user can combine different intertextual elements that refer to the same entity. When a text element is marked, it automatically generates its own entity and gets added to the panel under its respective category. At this point, the user can edit the Entity label or drag and drop one mention or entity with another that is truly representative of that mention. Updating an entity will synchronously update all the references in the HTML document to match the latest changes. The user can also click on the individual items to view the list of occurrences within the text. By clicking once on the row corresponding to an occurrence, one can automatically navigate to its location within the text through the \verb|about| attribute in the aforementioned \verb|span| element. In addition, each occurrence is itself draggable, allowing the integration with another mention, or its deletion, if necessary.

\begin{figure}[ht]
  \centering
  \includegraphics[width=\linewidth]{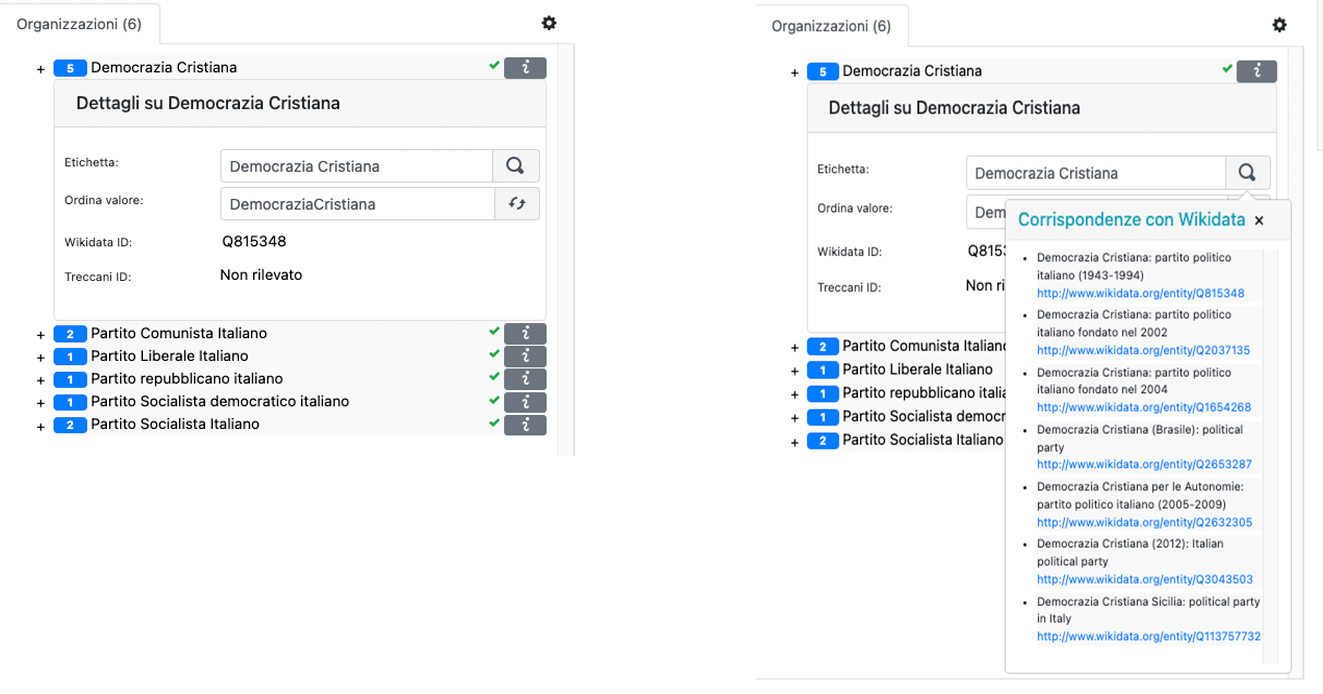}
  \caption{Wikidata entities referring to the organization "Democrazia Cristiana"}
  \Description{Wikidata entities referring to the organization "Democrazia Cristiana"}
  \label{wikidata}
\end{figure}

Each entity shows the total number of marked instances for that element and an info button to show details and edit the entity data. The checkbox indicates whether a Wikidata identifier is linked to the element. From the Entity details panel, the user can:

\begin{itemize}
    \item Update the label that identifies the element, in order to facilitate the query to the Wikidata knowledge graph, thus creating a relation with the respective Wikidata record, if present (Figure \ref{wikidata}).
    \item Update the value by which it is sorted within the entity list;
    \item the Wikidata identifier associated with the element;
    \item the Treccani identifier associated with that entity, if available. This identifier is stored in the Wikidata entity details. It is possible that Wikidata has not been populated with this value; in this case, the field is populated with the entry “Not Detected.”
\end{itemize}    

\subsection{Metadata insertion}
\begin{figure}[ht]
  \centering
  \includegraphics[width=\linewidth]{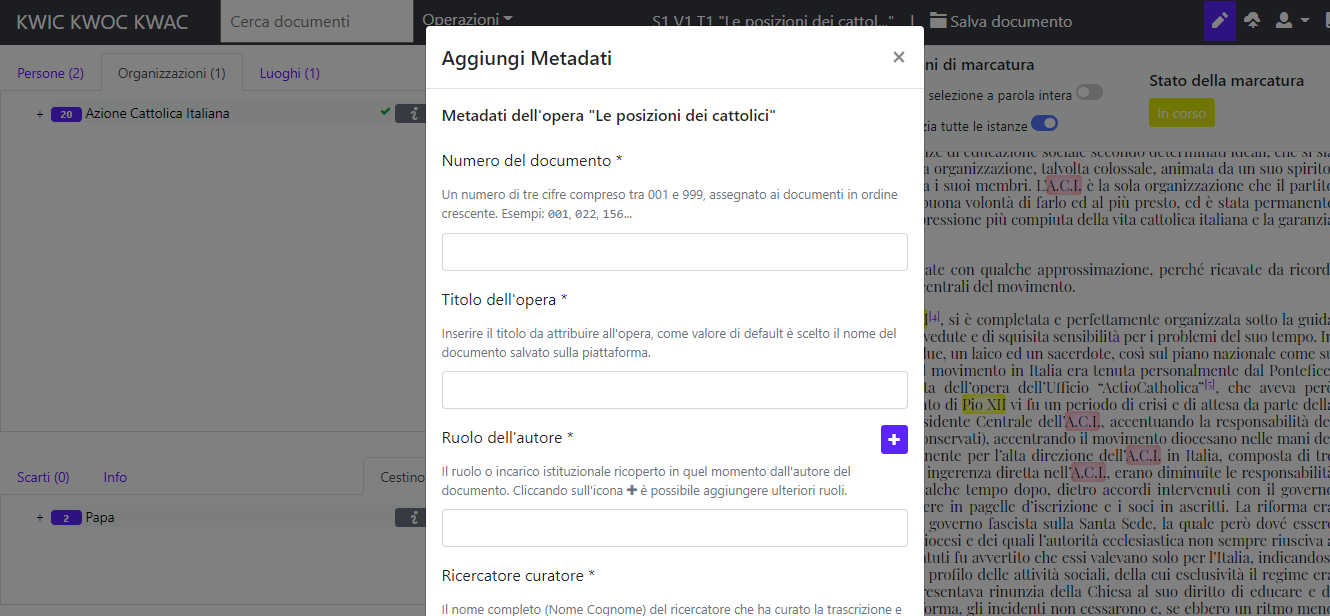}
  \caption{Metadata insertion}
  \Description{Metadata insertion}
  \label{metadata}
\end{figure}

Metadata are saved in a external MongoDB database using the \verb|mongoose| Javascript library. Along with the metadata, user info is also stored in this database. Within KwicKK's interface (Figure \ref{metadata}), the user can access a modal containing a form with the following fields, each used to enter one or more metadata values:

\begin{itemize}
    \item Document number: a three-digit number from 001 to 999, to be entered in ascending chronological order (e.g., the first document will have 001, the second 002, etc.);
    \item Author role: the role held by Moro at the time when the document was written;
    \item Researcher curator: the researcher name;
    \item Abstract: the description of the document prepared by the researcher;
    \item Document type: one or more categories the document belongs to;
    \item Document subject: one or more categories the subject of the document belongs to;
    \item Document status: indicates whether the document has been published/edited, or is unpublished/unedited;
    \item Bibliographic reference/Archival signature: one or more indication of the document's provenance (editorial source reference if published; archival signature if unpublished);
    \item Event place: name of the place where the event described in the document occurred;
    \item Date of event: date (day-month-year or year only) when the event described in the document occurred;
    \item Additional notes: a free-text field to store unstructured textual data containing supplemental information that could not be captured in the other fields.    
\end{itemize}

\section{Conclusions}

In this paper, we presented KwicKwocKwac 1.0 (KwicKK), a web-based annotation tool designed to facilitate the semi-automatic markup and enrichment of digital texts, particularly in the humanities domain. By providing intuitive functionalities for entity recognition, metadata insertion, and KWIC-based indexing of mentioned entities, KwicKK provides means that lower the barrier to entry for scholars and researchers, regardless of their technical proficiency. Its integration with RDFa and Linked Open Data, particularly through Wikidata, enhances the interconnectivity of scholarly resources, enabling users to link their data with existing authoritative sources. 

Future developments may focus on conducting further user experience research, allowing in-app text editing, integrating additional data sources, and expanding collaborative features. 

\bibliographystyle{alpha}
\bibliography{bibliography}

\end{document}